# Validation of constant mean free path and relaxation time approximations for metal resistivity: Explicit treatment of electron-phonon interactions

Subeen Lim[1], Yumi Kim[1], Gyungho Maeng[1], and Yeonghun Lee[1,2,3]*

[1] Department of Electronics Engineering, Incheon National University, Incheon 22012, Republic of Korea

[2] Department of Intelligent Semiconductor Engineering, Incheon National University, Incheon 22012, Republic of Korea

[3] Research Institute for Engineering and Technology, Incheon National University, Incheon 22012, Republic of Korea

**Corresponding author**

* Email: y.lee@inu.ac.kr

**Abstract**

The figure of merit $\rho\lambda$—the product of resistivity and mean free path (MFP)—evaluated from first-principles calculations, is widely adopted to screen promising interconnect metals with high electrical conductivity at ultranarrow dimensions. However, the $\rho\lambda$ has been calculated without addressing the validity of the assumption that the MFP is independent of the wavevector **k**. Here, we assess the validity of the constant MFP approximation, by estimating the **k**-dependent MFPs for (an)isotropic elemental metals, with explicit treatment of electron-phonon interactions. Additionally, we verify the validity of the constant relaxation time approximation (CRTA) for resistivity calculations. We show that both the constant MFP approximation and CRTA are reasonable even for highly anisotropic Fermi surfaces. Our results support the practical use of those approximations in transport studies, where explicit electron-phonon calculations are not feasible.

# 1. Introduction

As the ongoing miniaturization of electronic devices requires much narrower interconnects with low resistivity, understanding of the resistivity scaling properties of metals at narrow sizes has been a crucial topic over several decades [1–8]. The main property, in this matter, is significant increase of resistivity at ultranarrow dimensions [9,10]. This increase of resistivity in downscaled interconnects leads to a substantial RC delay and power dissipation in high-density integrated circuits (ICs) [11]. Indeed, resistivity scaling properties are characterized through a figure of merit known as $\rho\lambda$, which gauges the performance of ultranarrow interconnects [4]. However, obtaining a reliable estimate of $\rho\lambda$ in highly anisotropic or two-dimensional conductors requires descriptors that incorporate the anisotropy and directionality of physical quantities, such as group velocities [12]. A rigorous calculation of $\rho\lambda$ is challenging, as explicit calculations of electron-phonon interaction are computationally demanding. To circumvent this difficulty, the constant mean free path (MFP) approximation—assuming the MFP is uniform, **k**-independent—has been extensively applied in the screening of promising next-generation interconnect materials [13–18]. An alternative avoiding this difficulty is to employ deformation potential, which depends on long-wavelength acoustic phonons and band edge deformation potential [19]; it is inadequate and does not generalize to complex or metal systems [20–22]. Additionally, in the same consideration, the constant relaxation time approximation (CRTA) is often employed to analyze intrinsic electrical resistivity [23–25] and to evaluate the performance of thermoelectric devices [26,27]. Yet, it is noteworthy that these simplification neglecting **k-** and temperature-dependence can lead to an overestimation of the true figure of merit [28,29]. Therefore, to ensure reliable predictions of the transport properties, we assess the validity of the constant MFP approximation and CRTA by explicitly treating electron-phonon interactions.

In this work, we estimate $\rho\lambda$ and $\rho\tau$—the product of resistivity and mean free time—for (an)isotropic elemental metals, comparing cases with and without consideration of **k**-dependent MFP and relaxation time. We then discuss the validity of the constant MFP approximation and CRTA by analyzing the distributions of MFPs and relaxation times. Furthermore, the Pearson correlation coefficient is employed to quantify differences in the transport coefficient factors when **k**-dependence is considered versus when it is not. Our work elucidates that transport properties obtained solely from first-principles electronic band calculations, without treatment of electron-phonon interaction, remain sufficiently robust despite assuming **k**-independent MFPs and relaxation times.

## 2. Methods

Combining the Boltzmann transport equation and relaxation time approximation [30], the electrical conductivity tensor $\sigma_{\alpha\beta}$ (the inverse of electrical resistivity tensor $\rho_{\alpha\beta}$) can be calculated as

$$\sigma_{\alpha\beta} = [\rho_{\alpha\beta}]^{-1} = \frac{e^2}{(2\pi)^3} \sum_n \int_{BZ} v_n^\alpha(\mathbf{k}) v_n^\beta(\mathbf{k}) \tau_n(\mathbf{k}) \times \left.\frac{\partial f^0(\epsilon)}{\partial \epsilon}\right|_{\epsilon = E_n(\mathbf{k})} d^3\mathbf{k}, \quad (1)$$

where n is the band index, $\mathbf{k}$ is the electronic wave vector, $\alpha$ and $\beta$ indicate Cartesian directions, $e$ is the magnitude of the electron charge, $\mathbf{v}_n(\mathbf{k}) = \nabla_\mathbf{k} E_n(\mathbf{k})/\hbar$ is the group velocity, $\tau_n(\mathbf{k})$ is electron relaxation time which can be calculated under electron-phonon scattering matrix, $f^0(\epsilon)$ is the Fermi-Dirac distribution function, and $\epsilon$ is the eigenenergy at state $\mathbf{k}$.

Based on the electrical conductivity tensor in equation (1), we take into account the constant MFP approximation, $\lambda_n(\mathbf{k}) = |\mathbf{v}_n(\mathbf{k})|\tau_n(\mathbf{k}) = \lambda$, which naturally yields the following expression:

$$\frac{1}{\rho_{\alpha\beta}\lambda} = \frac{e^2}{(2\pi)^3} \sum_n \int_{BZ} \frac{v_n^\alpha(\mathbf{k}) v_n^\beta(\mathbf{k})}{|\mathbf{v}_n(\mathbf{k})|} \times \left.\frac{\partial f^0(\epsilon)}{\partial \epsilon}\right|_{\epsilon = E_n(\mathbf{k})} d^3\mathbf{k}. \quad (2)$$

For CRTA, the relaxation time becomes $\tau_n(\mathbf{k}) = \tau$ and therefore

$$\frac{1}{\rho_{\alpha\beta}\tau} = \frac{e^2}{(2\pi)^3} \sum_n \int_{BZ} v_n^\alpha(\mathbf{k}) v_n^\beta(\mathbf{k}) \times \left.\frac{\partial f^0(\epsilon)}{\partial \epsilon}\right|_{\epsilon = E_n(\mathbf{k})} d^3\mathbf{k}. \quad (3)$$

The $\mathbf{k}$-dependent relaxation time is calculated by considering the electron-phonon scattering probability as

$$\tau_n(\mathbf{k})^{-1} = \frac{2\pi}{N_q \hbar} \sum_{m,\nu\mathbf{q}} W_{n\mathbf{k},m\mathbf{k}+\mathbf{q}}^{\nu\mathbf{q}}, \quad (4)$$

where $N_q$ is the number of **q**-points, $W^{v\mathbf{q}}_{n\mathbf{k},m\mathbf{k}+\mathbf{q}}$ is the scattering probability between the electronic states $(n, \mathbf{k})$ and $(m, \mathbf{k}+\mathbf{q})$ with respect to the phonon mode $(v, \mathbf{q})$ denoted by mode index $v$ and wavevector **q**. The scattering probability can be expressed as [31]

$$W^{v\mathbf{q}}_{n\mathbf{k},m\mathbf{k}+\mathbf{q}} = \frac{2\pi}{\hbar} |g_{mnv}(\mathbf{k},\mathbf{q})|^2 [\delta(\epsilon_{n\mathbf{k}} - \hbar\omega_{v\mathbf{q}} - \epsilon_{m\mathbf{k}+\mathbf{q}})(1 + N^0_{v\mathbf{q}} - f^0_{m\mathbf{k}+\mathbf{q}})$$
$$+ \delta(\epsilon_{n\mathbf{k}} + \hbar\omega_{v\mathbf{q}} - \epsilon_{m\mathbf{k}+\mathbf{q}})(N^0_{v\mathbf{q}} + f^0_{m\mathbf{k}+\mathbf{q}})], \tag{5}$$

where $g_{mnv}(\mathbf{k},\mathbf{q})$ is electron-phonon matrix element, $\omega_{v\mathbf{q}}$ is the phonon frequency, $\delta$ is Dirac delta function and $N^0_{v\mathbf{q}}$ is the Bose-Einstein phonon occupations. This $W^{v\mathbf{q}}_{n\mathbf{k},m\mathbf{k}+\mathbf{q}}$ includes both phonon emission and absorption processes.

To evaluate the $\lambda_n(\mathbf{k})$ that contributes to conduction, the Fermi-Dirac weighted average MFP, $\langle\lambda\rangle$ is calculated as

$$\langle\lambda\rangle = \frac{\sum_n \int_{BZ} \lambda_n(\mathbf{k}) \times \left.\frac{\partial f^0(\epsilon)}{\partial \epsilon}\right|_{\epsilon=E_n(\mathbf{k})} d^3\mathbf{k}}{\sum_n \int_{BZ} \left.\frac{\partial f^0(\epsilon)}{\partial \epsilon}\right|_{\epsilon=E_n(\mathbf{k})} d^3\mathbf{k}}. \tag{6}$$

Same averaging manner applies to $\langle\tau\rangle$,

$$\langle\tau\rangle = \frac{\sum_n \int_{BZ} \tau_n(\mathbf{k}) \times \left.\frac{\partial f^0(\epsilon)}{\partial \epsilon}\right|_{\epsilon=E_n(\mathbf{k})} d^3\mathbf{k}}{\sum_n \int_{BZ} \left.\frac{\partial f^0(\epsilon)}{\partial \epsilon}\right|_{\epsilon=E_n(\mathbf{k})} d^3\mathbf{k}}. \tag{7}$$

The $\rho_{\alpha\beta}\langle\lambda\rangle$ and $\rho_{\alpha\beta}\langle\tau\rangle$ were calculated from the equations (1), (6) and (7).

The first-principles calculations based on density function theory (DFT) [32,33] and density function perturbation theory (DFPT) [34] are performed using QUANTUM ESPRESSO software [35,36], along with the Perdew-Burke-Ernzerhof (PBE) generalized gradient approximation (GGA)[37] and the norm-conserving pseudopotentials [38], available at the PseudoDojo library [39]. The energy cutoff is set to 102 Ry, corresponding to the highest cutoff value suggested in the Pseudo-dojo table among the elements. All crystal structures are fully relaxed until the residual atomic forces are less than 26 meV/Å, using a **k**-mesh of 15×15×15. Our DFT optimized lattice constants exhibit good agreement with experimental values, as shown in Table S1 of the Supplementary Material (SM). **k**-meshs of 24×24×24 for the cubic systems and 18×18×12 for the hexagonal systems are used, and energy criterion is set to $10^{-12}$ Ry for self-consistent field calculations following the structure relaxations. Spin polarization calculations are conducted for Co. The electron-phonon matrix elements are first calculated on the coarse **k** and **q** mesh ($8 \times 8 \times 8$). Then they are interpolated by the maximally localized Wannier functions (MLWFs) [40]. The transport properties are computed on dense **k** and **q** mesh ($80 \times 80 \times 80$) using PERTURBO code [31]. Additionally, the scattering rate is obtained using 100,000 uniformly distributed random **q**-points.

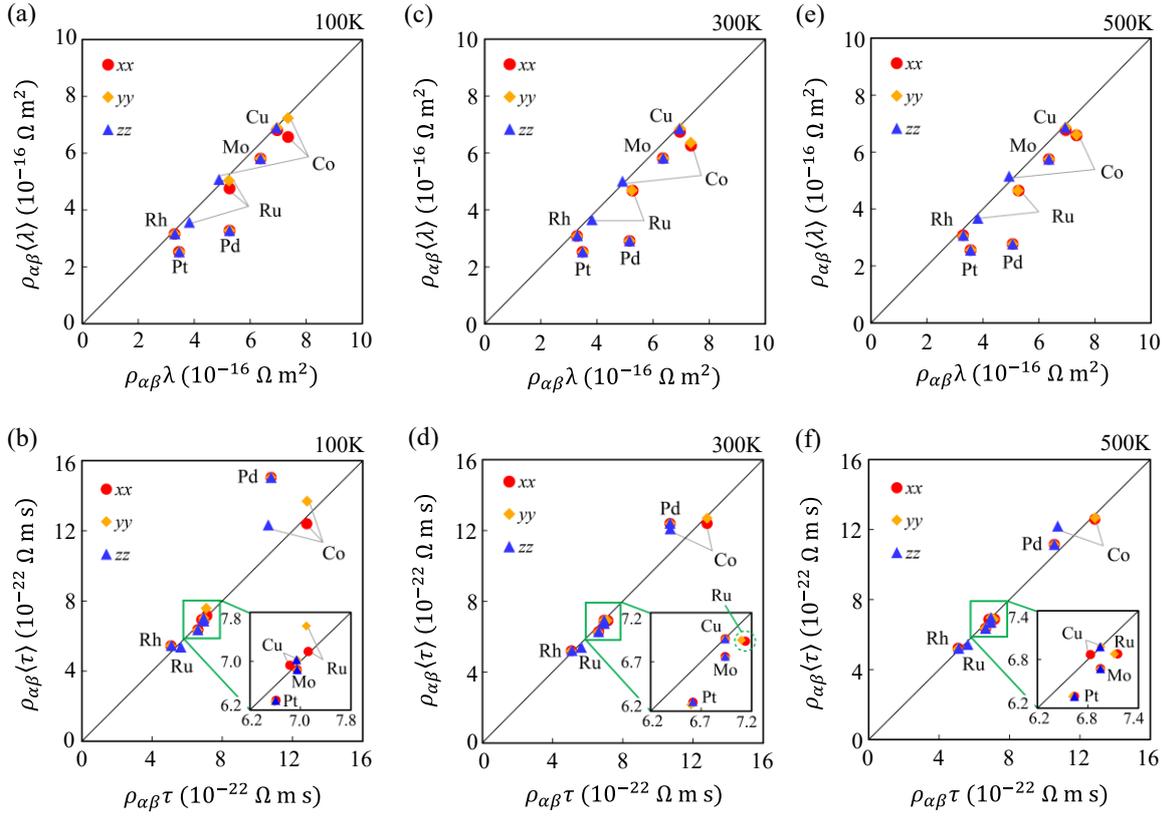

**Figure 1.** Comparison of $\rho_{\alpha\beta}\lambda$ and $\rho_{\alpha\beta}\langle\lambda\rangle$ considering **k**-dependent MFPs, and comparison of $\rho_{\alpha\beta}\tau$ and $\rho_{\alpha\beta}\langle\tau\rangle$ considering **k**-dependent relaxation times for elemental metals with different temperatures. Circles, diamonds, and triangles denote *xx*, *yy*, and *zz* tensor elements, respectively.

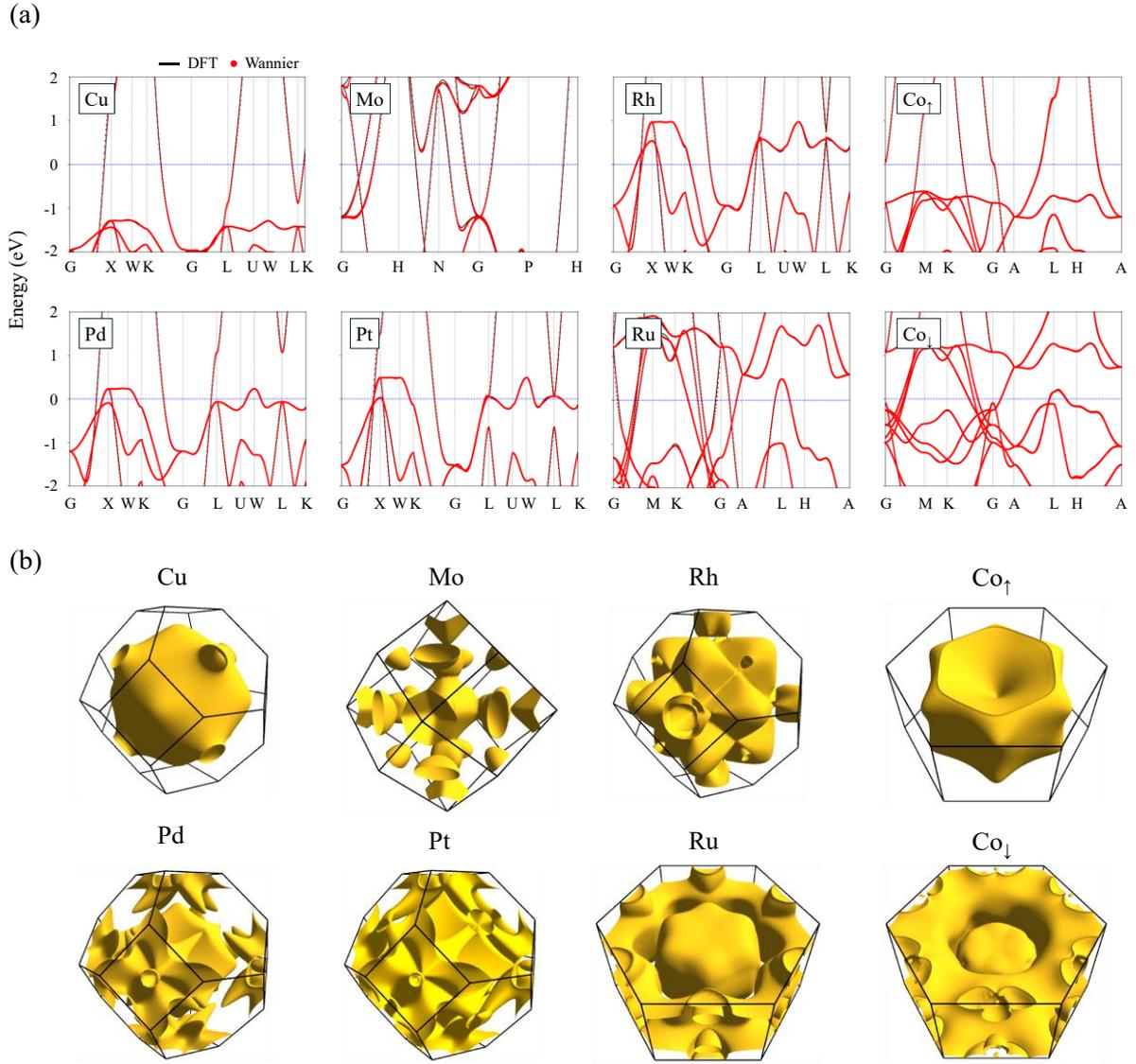

**Figure 2.** (a) Calculated band structures of elemental metals. Black solid lines and red dots represent the DFT and the Wannier interpolated band structures. Blue dashed line indicates the Fermi level. (b) Calculated Fermi surfaces of elemental metals.

## 3. Results and discussion

To evaluate the impact of the **k**-dependent MFPs and relaxation times on the transport properties, we compared $\rho_{\alpha\beta}\langle\lambda\rangle$ and $\rho_{\alpha\beta}\langle\tau\rangle$ at different temperatures (100, 300, and 500 K) with the values obtained under the assumption of **k**-independent MFPs and relaxation times (see figure 1). We selected Cu as a standard reference, along with promising metals for interconnect such as platinum-group metals (Ru, Rh, Pd, and Pt) [15,41,42], Mo [43], and Co [44]. As shown in figure 1, both $\rho_{\alpha\beta}\lambda$ and $\rho_{\alpha\beta}\tau$, which employ the constant MFP approximation and the CRTA, respectively, remain in good agreement with $\rho_{\alpha\beta}\langle\lambda\rangle$ and $\rho_{\alpha\beta}\langle\tau\rangle$ for all directions at a wide range of temperatures. It should be noted that, however, that Pd exhibits deviations within the CRTA at 100 K, indicating that the use of the CRTA at low temperatures requires particular caution. In the following analysis, we focus on 300 K, as the transport properties show negligible temperature dependence. To estimate the errors induced by **k**-independent approximation, we calculated the root-mean-square error (RMSE) for $\rho_{\alpha\beta}\langle\lambda\rangle$ and $\rho_{\alpha\beta}\langle\tau\rangle$. Figure 1(c) yielded an RMSE of $1.41\times10^{-16}\Omega\,\text{m}^2$, while Figure 1(d) produced a lower RMSE of $1.02\times10^{-22}\Omega\,\text{m}\,\text{s}$. It seems that the transport properties adopting CRTA are robust enough, while those obtained within the constant MFP approximation do not achieve the same level of accuracy. However, for $\rho_{\alpha\beta}\lambda$, they do not lead to significant errors when $\rho_{\alpha\beta}\lambda$ is used in the preliminary screening stage for selecting candidate materials for ultranarrow interconnects, although Pt and Pd exhibit minor deviations [see figure 1(c)]. Figure 2 presents the band structures and the Fermi surfaces for Cu, Mo, Pd, Pt, Rh, Ru, and Co, with Brillouin zones depicted in black lines. As can be seen in figure 2(b), the Fermi surface of Cu is spherical and isotropic, which means that the transport properties are not sensitive to the constant MFP approximation and CRTA. Therefore, it is not surprising that Cu shows negligible differences in the value of the transport coefficient factors, regardless of whether **k**-dependent MFPs and relaxation times are considered. On the other hand, the Fermi surfaces of the other metals are non-spherical and highly anisotropic (see figure 2(b)), which can lead to a pronounced sensitivity to the **k**-dependent MFPs and relaxation times. Despite the non-spherical and anisotropic nature of the Fermi surfaces in the other metals, the calculated $\rho_{\alpha\beta}\langle\lambda\rangle$ and $\rho_{\alpha\beta}\langle\tau\rangle$ do not exhibit significant deviation with the **k**-independent results.

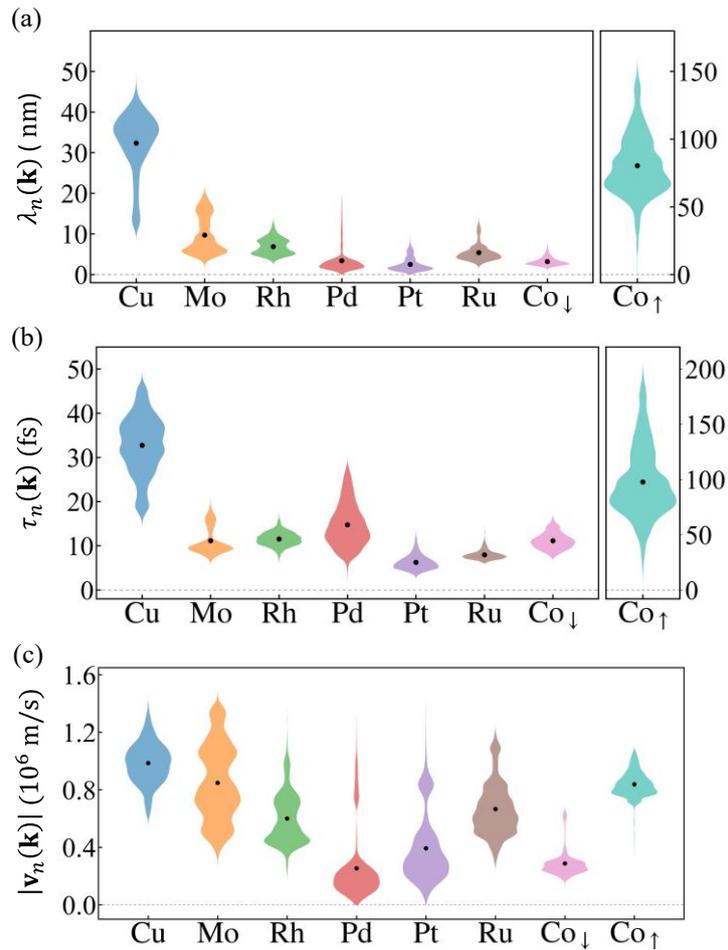

**Figure 3.** Distributions of (a) MFPs, (b) relaxation times, and (c) group velocity magnitudes for each elemental metal at 300 K. All the values are evaluated for states within $E_\text{F} \pm 0.1\text{eV}$.

Black dots indicate the mean values.

**Table 1.** Calculated CVs of $\lambda_n(\mathbf{k})$ and $\tau_n(\mathbf{k})$, and $|\mathbf{v}_n(\mathbf{k})|$ at 300 K. All the values are evaluated within $E_\mathrm{F} \pm 0.1\mathrm{eV}$.

| | Materials | | | | | | | |
|---|---|---|---|---|---|---|---|---|
| CV | Cu | Mo | Rh | Pd | Pt | Ru | Co↓ | Co↑ |
| $\lambda_n(\mathbf{k})$ | 0.23 | 0.47 | 0.29 | 0.91 | 0.68 | 0.38 | 0.28 | 0.32 |
| $\tau_n(\mathbf{k})$ | 0.20 | 0.27 | 0.14 | 0.21 | 0.27 | 0.13 | 0.16 | 0.33 |
| $|\mathbf{v}_n(\mathbf{k})|$ | 0.14 | 0.30 | 0.29 | 0.92 | 0.59 | 0.27 | 0.32 | 0.11 |

To further assess the validity of the constant MFP approximation and CRTA, we analyzed the distributions of MFPs, relaxation times, and group velocity magnitudes for different $\mathbf{k}$ (see figure 3). Although the MFPs exhibit a substantial $\mathbf{k}$-dependence shown in figure 3(a), the figures of merit $\rho_{\alpha\beta}\lambda$ obtained from the constant MFP approximation generally show good agreement with those from $\mathbf{k}$-dependent MFPs [see figure 1(c)]. This observation motivates a more detailed investigation of the MFP distribution. We find that the more distorted the mean of the $\mathbf{k}$-dependent MFP distribution becomes, the more likely it is that the $\mathbf{k}$-independent MFP approximation becomes invalid. In figure 3(a), the mean values generally do not significantly deviate from the peak of the distribution of $\lambda_n(\mathbf{k})$, even though some metals represent anisotropic Fermi surfaces. For anisotropic (hcp) metals, the distribution of MFPs does not broaden, and the mean lies close to the peak. It is linked to the result that the constant MFP approximation is robust enough. However, Pd and Pt showed a slight difference between $\rho_{\alpha\beta}\langle\lambda\rangle$ and $\rho_{\alpha\beta}\lambda$ [see figure 1(c)]. To quantitatively estimate the distributions of $\lambda_n(\mathbf{k})$, we compared the coefficients of variation (CVs), $\mathrm{CV} = \sigma/\mu$, where $\sigma$ is the standard deviation and $\mu$ is the mean of the distribution. Pd and Pt show the highest CVs in their MFP distributions, 0.91 and 0.68, respectively, as listed in Table 1. It indicates that Pd and Pt possess the most non-uniform MFP distributions among the metals, which in turn lead to anomalous deviation in $\rho\lambda$ for these two elements. For all metals, the distributions of $\tau_n(\mathbf{k})$ show more symmetric shape than those of $\lambda_n(\mathbf{k})$, which implies that the mean is closely aligned with the distribution peak, as illustrated in figure 3(b). In addition, $\tau_n(\mathbf{k})$ exhibits a low CV for all metals, implying a relatively uniform distribution of $\tau_n(\mathbf{k})$, in contrast to the result of $\lambda_n(\mathbf{k})$, as shown in Table 1. This suggests that the CRTA is less sensitive to $\mathbf{k}$-dependence than the constant MFP approximation. Afterwards the differences in $\rho_{\alpha\beta}\lambda$ and $\rho_{\alpha\beta}\tau$ due to the $\mathbf{k}$-dependence will be further quantified through correlation analysis in figures 4 and 5. For group velocity magnitude $|\mathbf{v}_n(\mathbf{k})|$, most metals—except for Cu—appear to exhibit anisotropic distributions, especially for Pd. In Table 1, we can expect that Pd and Pt have non-uniform distributions of $|\mathbf{v}_n(\mathbf{k})|$ due to the CV values of 0.92 and 0.59, respectively. Based on the relationship, $\lambda_n(\mathbf{k}) = |\mathbf{v}_n(\mathbf{k})|\tau_n(\mathbf{k})$, we found that the non-uniform MFPs are induced by the non-uniform $|\mathbf{v}_n(\mathbf{k})|$. Consequently, as distributions of $\tau_n(\mathbf{k})$ generally show uniformity, the complex and highly anisotropic Fermi surface can lead to anisotropic distribution of MFPs, making anomalous outcome under the constant MFP approximation.

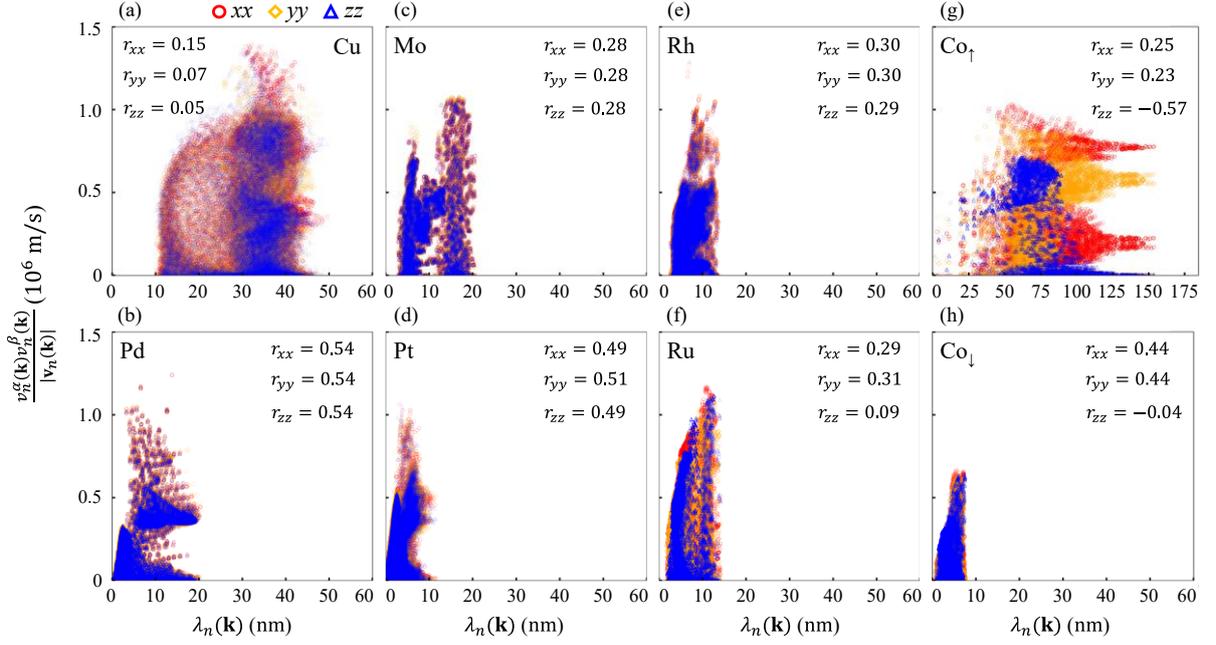

**Figure 4.** Relationship between $v_n^\alpha(\mathbf{k})v_n^\beta(\mathbf{k})/|\mathbf{v}_n(\mathbf{k})|$ and $\lambda_n(\mathbf{k})$ for each elemental metal at 300 K. Pearson correlation coefficients are provided as $r_{\alpha\beta}$. Data are evaluated for states within $E_\mathrm{F} \pm 0.1\mathrm{eV}$. Circles, diamonds, and triangles denote *xx*, *yy*, and *zz* components, respectively.

To quantitatively explain the deviations in $\rho_{\alpha\beta}\lambda$ and $\rho_{\alpha\beta}\tau$ arising from $\mathbf{k}$-dependent MFPs and relaxation times, we performed correlation analyses. In equation (2), the MFP, originally inside the $\mathbf{k}$-integral on the right-hand side, is treated as a constant and brought to the left-hand side, implying a possible correlation with group velocity. This suggests that if there exists a strong correlation between the MFP and group velocity, the constant MFP approximation may lead to an inaccurate evaluation of the figure of merit. Similarly, equation (3) suggests that a strong correlation between relaxation time and group velocity can result in deviations of $\rho_{\alpha\beta}\tau$ under the CRTA. To gain intuition, we first performed a correlation analysis between $|\mathbf{v}_n(\mathbf{k})|$ and $\lambda_n(\mathbf{k})$, as shown in figure S1 of the SM. Most elemental metals exhibit large Pearson correlation coefficients, a widely used measure of linear dependence between continuous variables, indicating strong correlations. However, these strong correlations do not always translate into significant deviations of $\rho_{\alpha\beta}\lambda$ from $\rho_{\alpha\beta}\langle\lambda\rangle$ as can be seen in figure 1(c). To further investigate these discrepancy, we conducted a correlation analysis between $v_n^\alpha(\mathbf{k})v_n^\beta(\mathbf{k})/|\mathbf{v}_n(\mathbf{k})|$—inside the $\mathbf{k}$-integral in equation (2)—and the $\lambda_n(\mathbf{k})$ (see figure 4). In figures 4(b) and 4(d), Pd and Pt exhibit large Pearson correlation coefficients around 0.5 compared to the other metals, consistent with the deviations in the figure of merit caused by $\mathbf{k}$-dependence, as observed in figure 1(c).

Additionally, in figures 4(g) and 4(h), spin-up Co exhibits strong negative correlation in the *zz* component with Pearson correlation values of -0.57, while spin-down Co presents strong positive correlation in the *xx* and *yy* components with Pearson correlation values of 0.44. Notwithstanding, as shown in figure 1(c), Co shows only slight difference between $\rho_{\alpha\beta}\lambda$ and $\rho_{\alpha\beta}\langle\lambda\rangle$ in all directions. It should be noted that within the spin polarized case, the two spin channels are treated as independent channels that conduct in parallel, which can be written as:

$$\rho_{\alpha\beta}\langle\lambda\rangle = \frac{\rho_{\alpha\beta}^\uparrow\langle\lambda^\uparrow\rangle \times \rho_{\alpha\beta}^\downarrow\langle\lambda^\downarrow\rangle}{\rho_{\alpha\beta}^\uparrow\langle\lambda^\uparrow\rangle + \rho_{\alpha\beta}^\downarrow\langle\lambda^\downarrow\rangle}. \qquad (8)$$

Therefore, due to the half harmonic averaging in equation (8), the Pearson correlation coefficient of Co becomes reduced for each direction, despite the strong individual correlations of spin-up and spin-down channels. Specifically, since the harmonic mean is dominated by smaller values, the out-of-plane Pearson correlation coefficient becomes significantly lower than the in-plane components (e.g., -0.04 for spin-down Co in the *zz* component). As a result, the difference between $\rho_{\alpha\beta}\lambda$ and $\rho_{\alpha\beta}\langle\lambda\rangle$ for the Co along the *zz* component becomes much smaller, as listed in figure 1(c). In terms of CRTA, we examined the relationship between the $|\mathbf{v}_n(\mathbf{k})|$ and $\tau_n(\mathbf{k})$, as illustrated in figure S2 of the SM. We observed that the Pearson correlation coefficients between $|\mathbf{v}_n(\mathbf{k})|$ and $\tau_n(\mathbf{k})$ are small. Nonetheless, Ru exhibits strong correlation, which is not matched with the result in figure 1(d). In figure 5, we therefore calculated the Pearson correlation coefficient between $v_n^\alpha(\mathbf{k})v_n^\beta(\mathbf{k})$ and $\tau_n(\mathbf{k})$ to accurately assess the validity of the CRTA. The results show weak correlations across most elemental metals, which is in good agreement with the trend observed in figure 1(d). Particularly, both spin-up and spin-down Co exhibit relatively large out-of-plane Pearson correlation values. However, the harmonic averaging again leads to a smaller correlation. As a result, as shown in figure 1(d), the difference of Co between $\rho_{\alpha\beta}\tau$ and $\rho_{\alpha\beta}\langle\tau\rangle$ becomes less pronounced. For all metals we examined, $v_n^\alpha(\mathbf{k})v_n^\beta(\mathbf{k})$ and $\tau_n(\mathbf{k})$ exhibit only weak correlations. This result supports the conclusion that CRTA is a reasonable and highly effective approach for analyzing intrinsic electrical resistivity even in the presence of a complex and anisotropic Fermi surface.

The strong correlations in Pd and Pt, and Co (see Figure 4) can be explained by group velocity distributions. Understanding the correlation between the group velocity and the MFP requires first analyzing how the group velocity correlates with the relaxation time, which is the key component of the MFP. Figure 5 shows that correlation between group velocity and relaxation time is generally weak although Co exhibits strong negative correlation for *zz* component. The strong negative correlation in spin-up Co of *zz* component is induced by small group velocity along the *z* direction. Fermi surface of spin-up Co shows small curvature (see Figure 2(b)) along the out-of-plane direction, which indicates that $v_n^z(\mathbf{k})$ is smaller than $v_n^x(\mathbf{k})$ and $v_n^y(\mathbf{k})$. However, as shown in Figure S2(g), relatively large $|\mathbf{v}_n(\mathbf{k})|$ implies large in-plane components $v_n^x(\mathbf{k})$ and $v_n^y(\mathbf{k})$ when $v_n^z(\mathbf{k})$ is substantially small. The large $v_n^x(\mathbf{k})$ and $v_n^y(\mathbf{k})$ generally imply a reduced density of states, suggesting that fewer final states are available for scattering and, consequently that $\tau_n(\mathbf{k})$ can take on relatively large values, which produces the negative correlation between $v_n^z(\mathbf{k})$ and $\tau_n(\mathbf{k})$. Although Pd and Pt exhibit only weak correlations between the group velocity and the relaxation time, their correlations with the MFP become significantly stronger. To understand this behavior, we examined the correlation between $v_n^\alpha(\mathbf{k})v_n^\beta(\mathbf{k})/|\mathbf{v}_n(\mathbf{k})|$ and $|\mathbf{v}_n(\mathbf{k})|$, as can be seen in Figure S3, motivated by the fact that the MFP is defined as $\lambda_n(\mathbf{k}) = |\mathbf{v}_n(\mathbf{k})|\tau_n(\mathbf{k})$. Figure S3 shows strong positive correlation for Pd and Pt, which are consistent with the strong correlation between $v_n^\alpha(\mathbf{k})v_n^\beta(\mathbf{k})/|\mathbf{v}_n(\mathbf{k})|$ and $\lambda_n(\mathbf{k})$ discussed above. For the metals that exhibit weak correlations, we find that $|\mathbf{v}_n(\mathbf{k})|$ does not approach zero, indicating the absence of very small $|\mathbf{v}_n(\mathbf{k})|$, as can be seen in Figure S3. In contrast, when $|\mathbf{v}_n(\mathbf{k})|$ reaches values close to zero, as in Pd and Pt, the correlations between $v_n^\alpha(\mathbf{k})v_n^\beta(\mathbf{k})/|\mathbf{v}_n(\mathbf{k})|$ and $|\mathbf{v}_n(\mathbf{k})|$ become significantly enhanced. This, in turn, amplifies the correlation with $\lambda_n(\mathbf{k})$, which can deteriorate the validity of the constant MFP approximation. As shown in figure 2(a), Pd and Pt show flat band dispersions near the Fermi level, which imply group velocities approaching zero. Consequently, for metals such as Pd and Pt that possess flat bands at the Fermi level, applying the constant MFP approximation can lead to inaccurate results. This observation suggests that, when screening ultranarrow interconnect materials, systems expected to host flat bands near the Fermi level require a more accurate evaluation of $\rho\lambda$.

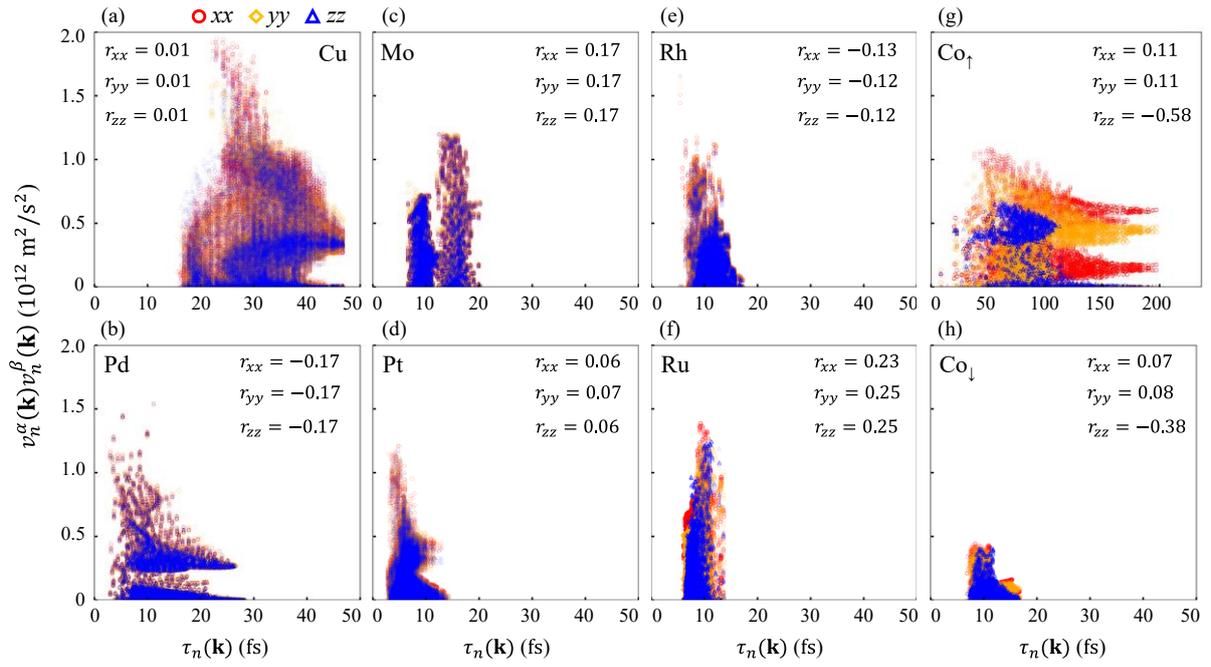

**Figure 5.** Relationship between $v_n^\alpha(\mathbf{k})v_n^\beta(\mathbf{k})$ and $\tau_n(\mathbf{k})$ for each elemental metal at 300 K. Pearson correlation coefficients are provided as $r_{\alpha\beta}$. Data are evaluated for states within $E_\mathrm{F} \pm 0.1\,\mathrm{eV}$. Circles, diamonds, and triangles denote *xx*, *yy* and *zz* components, respectively.

## 4. Conclusion

We investigated **k**-dependent MFPs and relaxation times by explicitly calculating electron-phonon interactions, in order to assess the validity of the constant MFP approximation and CRTA that are inherently adopted in $\rho_{\alpha\beta}\lambda$ and $\rho_{\alpha\beta}\tau$, respectively. Our results show that the constant MFP approximation and CRTA do not lead to significant deviations of transport properties, even in the presence of highly anisotropic Fermi surfaces. The MFP distribution and its correlation with the group velocity are investigated to identify additional factors responsible for the anomalous outcomes. Furthermore, our investigation of relaxation time and group velocity distributions explains that CRTA is robust and suitable for evaluating electrical resistivity. We believe that our work can provide strong validity for the use of both $\rho_{\alpha\beta}\lambda$ and $\rho_{\alpha\beta}\tau$, which are efficiently calculated from first principles electronic band data, and also support the practical applicability of those approximations to transport studies when explicit electron-phonon calculations are infeasible.

**Acknowledgment**

The authors thank Young-Rok Jang, Youngho Kang, Jeongwoo Kim, and Ki Hoon Lee for valuable discussions. This work was supported by K-CHIPS (Korea collaborative & High-tech Initiative for Prospective Semiconductor Research) (2410012153, RS-2025-02310666, 25073-15FC) funded by the Ministry of Trade, Industry & Energy (MOTIE, Korea).

# Supplementary Material

# Validation of constant mean free path and relaxation time approximations for metal resistivity: explicit treatment of electron-phonon interactions

Subeen Lim[1], Yumi Kim[1], Gyungho Maeng[1], and Yeonghun Lee[1,2,3]*

[1] Department of Electronics Engineering, Incheon National University, Incheon 22012, Republic of Korea

[2] Department of Intelligent Semiconductor Engineering, Incheon National University, Incheon 22012, Republic of Korea

[3] Research Institute for Engineering and Technology, Incheon National University, Incheon 22012, Republic of Korea

**Corresponding author**

* Email: y.lee@inu.ac.kr

## S1. Lattice constant

**Table S1.** Lattice structure and lattice constant of metals in our work, and experimental lattice constants for comparisons.

| Materials | Lattice structure | Lattice constant | |
|---|---|---|---|
| | | DFT (Å) | Exp. (Å) |
| Cu | FCC | 3.66 | 3.61[a] |
| Rh | FCC | 3.83 | 3.80[b] |
| Pd | FCC | 3.94 | 3.89[a] |
| Pt | FCC | 3.97 | 3.92[a] |
| Mo | BCC | 3.16 | 3.15[c] |
| Ru | HCP | $a=b=2.72$, $c=4.30$ | $a=b=2.71$[d], $c=4.28$[d] |
| Co | HCP | $a=b=2.50$, $c=3.96$ | $a=b=2.51$[a], $c=4.07$[a] |

[a]Reference [1].

[b]Reference [2].

[c]Reference [3].

[d]Reference [4].

## S2. Correlation analysis

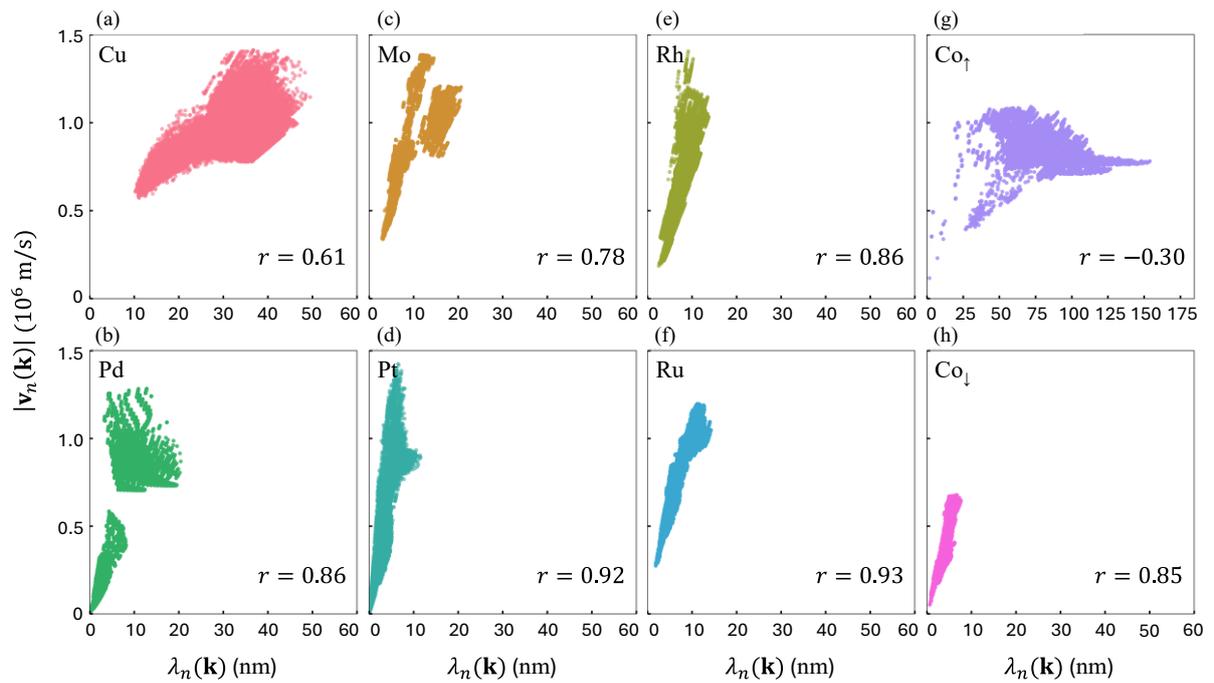

**Figure S1.** The group velocity magnitude is compared with the MFP for each elemental metal at 300 K. Pearson correlation coefficients are provided as $r$. Data are evaluated for states within $E_\mathrm{F} \pm 0.1\mathrm{eV}$.

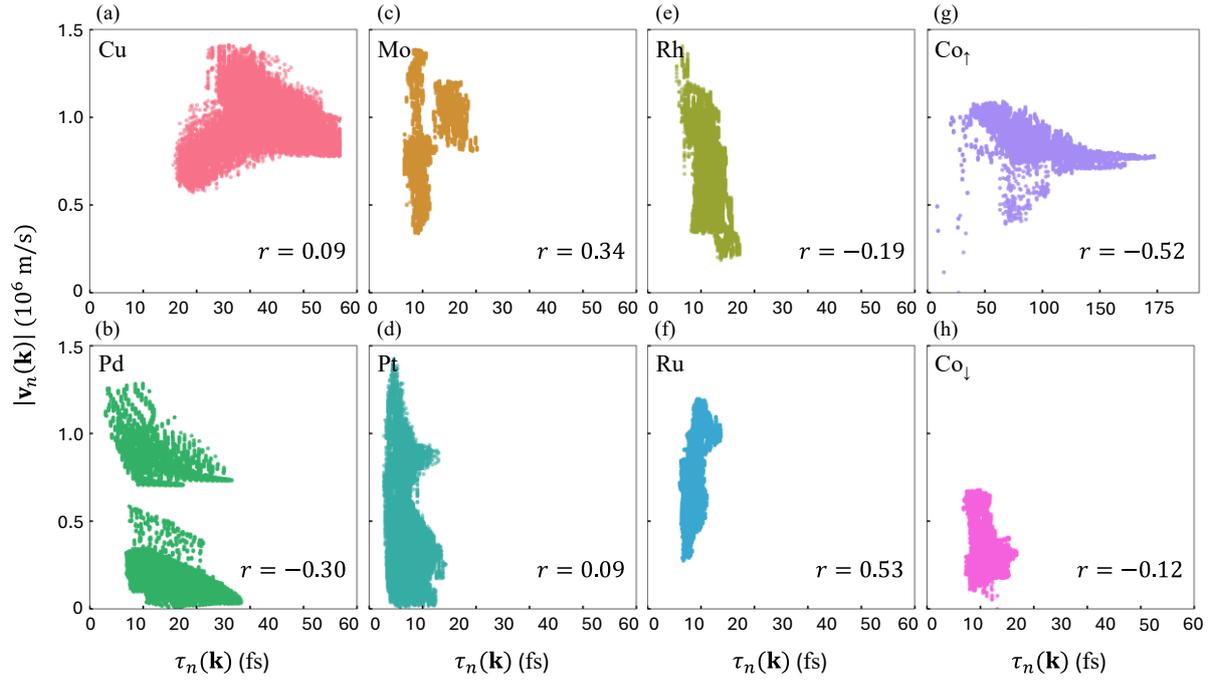

**Figure S2.** The group velocity magnitude is compared with the relaxation time for each elemental metal at 300 K. Pearson correlation coefficients are provided as $r$. Data are evaluated for states within $E_\mathrm{F} \pm 0.1\mathrm{eV}$.

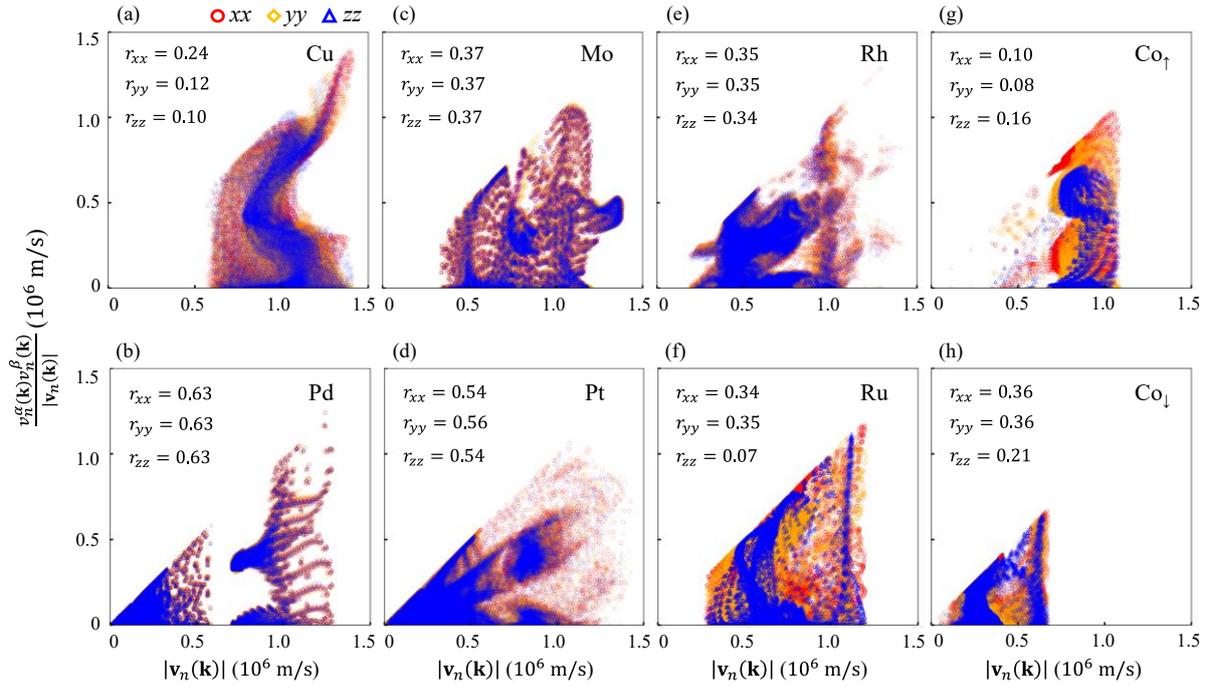

**Figure S3.** Relationship between $v_n^\alpha(\mathbf{k})v_n^\beta(\mathbf{k})/|\mathbf{v}_n(\mathbf{k})|$ and $|\mathbf{v}_n(\mathbf{k})|$ for each elemental metal at 300 K. Pearson correlation coefficients are provided as $r_{\alpha\beta}$. Data are evaluated for states within $E_\mathrm{F} \pm 0.1\,\mathrm{eV}$. Circles, diamonds, and triangles denote *xx*, *yy*, and *zz* components, respectively.